# DEEP BOREHOLE DISPOSAL OF NUCLEAR WASTE

[1]Patrick V. Brady, Geoffrey A. Freeze, Kristopher L. Kuhlman, Ernest L. Hardin, David C. Sassani, and Robert J. MacKinnon

[1]*Sandia National Laboratories, Albuquerque, New Mexico, USA 87185;*
*pvbrady@sandia.gov*

## Abstract

Radioactive waste disposal in deep boreholes may be more "ready" than disposal in mined geologic repositories since mankind has greater experience operating small deep holes – boreholes, than big shallow holes - mines. There are several thousand precedents for constructing > 2 km deep boreholes and several hundred precedents for disposing long-lived wastes in boreholes. Borehole disposal is likely to be faster, cheaper, and more flexible than mined disposal, while also offering greater long-term isolation. Isolation would rely on the great depth, water density gradients, and reducing conditions to prevent vertical movement of waste up the borehole.

## Keywords

## Introduction

Mankind has a long history drilling boreholes. The Chinese drilled oil wells in Sichuan Province to ~ 240 m depth in 347 AD using percussion drilling with bits attached to bamboo poles. Oil was used to evaporate brine to produce salt. By 1835 Chinese boreholes exceeded 1000 m depth. In the meantime, the Chinese had worked out pipelines; of split bamboo, well casing; of wooden logs, downhole cables; of bamboo twine, variable speed drilling, and "fishing" techniques for retrieving objects from the bottom of the well (Kuhn 2004). The number of deep boreholes ballooned in the 20th century with the search for oil, gas, and geothermal energy. Today the US alone drills over 40,000 boreholes annually for oil, gas, and geothermal purposes, and has been doing so for decades. Average borehole depths in recent years have been ~ 2 km; in the US there are 1.1 million conventional boreholes producing oil and gas.

The regulated disposal of long-lived, albeit non-nuclear, waste in boreholes is already happening in the US – in particular, approximately 800 Class I wells inject hazardous and non-hazardous industrial waste and municipal wastewater into geologic formations that are ~ 0.5 to 2 km deep (http://www2.epa.gov/uic/class-i-industrial-and-municipal-waste-disposal-wells). Forty-four Class I wells dispose of hazardous waste. The injection zone must be below drinking water aquifers, have sufficient hydraulic capacity to accept the hazardous waste, and be separated from drinking water aquifers by one or more impermeable, confining layers. There is a 10,000 year no-migration of wastes stipulation that must be demonstrated for each well. In short, there is already a successfully operating framework for disposing of long-lived non-nuclear waste in boreholes which might be applied to nuclear waste disposal (see below). This parallel has been noted by the US Environmental Protection Agency (Schultheisz 2015).

Recent articles in *Science* (Cornwall 2015) and *Nature* (Tollefson 2014) highlight the growing interest in deep borehole disposal of nuclear waste – and with good reason. Deep borehole disposal has many attractive features – robust isolation from the biosphere, low cost, speed of implementation, and modularity. In its simplest form (e.g. Arnold, Brady, Bauer, Herrick, Pye and Finger 2011), deep borehole disposal consists of drilling a large-diameter (up to 43 cm, or 17 in) borehole, or array of boreholes, into crystalline basement rock to a depth of about 5,000 m, emplacing waste packages in the lower ~2,000 m portion of the borehole, and sealing and plugging the upper portion of the borehole with a combination of bentonite, cement, and cement/crushed rock backfill. The concept is illustrated in Figure 1. Waste emplaced into a deep borehole would be several times deeper than in typical mined repositories. At these depths groundwater density stratification inhibits long-term vertical advection of groundwater and radionuclides, and geochemically reducing conditions lower radionuclide solubility and enhance sorption, all of which inhibits long-term transport. The temperature increase derived from radioactive decay of the waste, and the associated thermally-induced groundwater flow up the borehole lasts for only a few hundred years, after which slow diffusion is the predominant long-term transport mechanism. Borehole disposal relies on currently available commercial technology for drilling, emplacement, and sealing (Beswick, Gibb and Travis 2014). Deep borehole disposal is being examined as an option for radioactive waste disposal by researchers in the US, the United Kingdom, China, Germany, South Korea, and Australia.

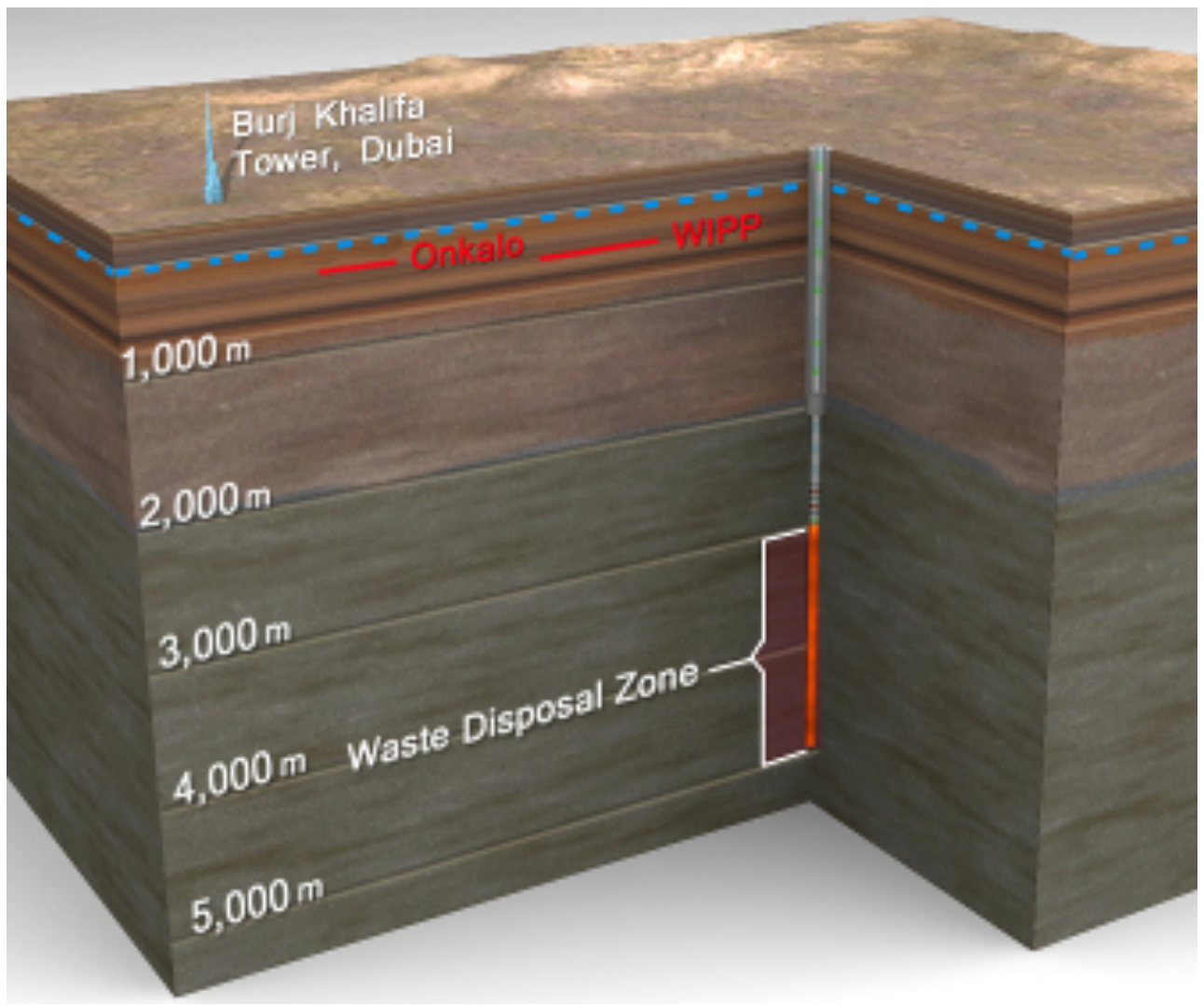

Figure 1. Generalized schematic of the deep borehole disposal concept (from Arnold, Vaughn, MacKinnon, Tillman, Nielson, Brady, Halsey and Altman 2012).

Borehole disposal of nuclear waste is expected to cost substantially less than traditional mined repositories. Brady et al. (2009) estimated borehole costs of ~ $40M for drilling, completion, and waste emplacement; the whole process for a single borehole would take less than 2 years (approximately 6 months for drilling and a year for emplacement, SNL 2015). More recent, and more extensive cost analysis of Bates (2015) examined spent fuel disposal and varied borehole depth, disposal zone length and borehole spacing in the calculation, while constraining post-closure dose. Bates (2015) established optimized disposal costs to be $45 - $191/kgHM (kg Heavy Metal). These "first-of-a-kind" costs should decrease with experience, but still are substantially lower than the $400/kgHM that was collected in the US nuclear waste fund.

Deep borehole disposal is modular, with construction and operational costs scaling linearly with waste inventory. This "pay as you go" approach avoids the large up-front investment of a mined geologic repository. If, for example, drilling shows that a deep borehole site is unsuitable, the drill rig is simply moved to another candidate site, and the process repeated until an acceptable site for disposal is found. The relative ease of finding crystalline basement at < 2 km depth means disposal could be decentralized to achieve a greater degree of geographic and political equity; it could also decrease transportation costs and risks. Deep borehole disposal should also be attractive to countries that have smaller radioactive waste inventories; the entire inventory could fit in a handful of boreholes. Borehole disposal for these countries' wastes would avoid the large expense of a mined geologic repository.

The presence of "old" saline and chemically reducing waters at depth is a key to the success of deep borehole disposal. The presence of old water at depth, which lost active contact with the surface hydrosphere hundreds of thousands of years ago is evidence that there is little driving force for upward water movement. The presence of dense, saline brines at depth is a barrier to buoyant upward movement of the water into overlying fresher water. Oxygen-poor, reducing conditions at depth slow the degradation of spent fuel and maintain many of the radionuclides in their lower valence states, lowering radionuclide solubility controls, and stabilizing most strongly-sorbed forms (Brady et al., 2009). Also, the great depth of deep borehole disposal decreases the number of surface effects that must be considered for long-term performance. These include groundwater infiltration, human intrusion, and effects from climate change including glaciation.

Hydrologic conditions at depth should also limit the upward advective transport of radionuclides in deep groundwater. Crystalline basement rocks generally have very low permeability. The only driving force for upward water flow could be expansion of water caused by heat from radioactive decay in the waste. The thermal heat pulse is an early feature that dissipates within approximately the first few hundred years after emplacement, or sooner depending on the waste type. Borehole seals of cement and/or bentonite would limit fluid movement during the early thermal pulse, and possibly for longer periods of time.

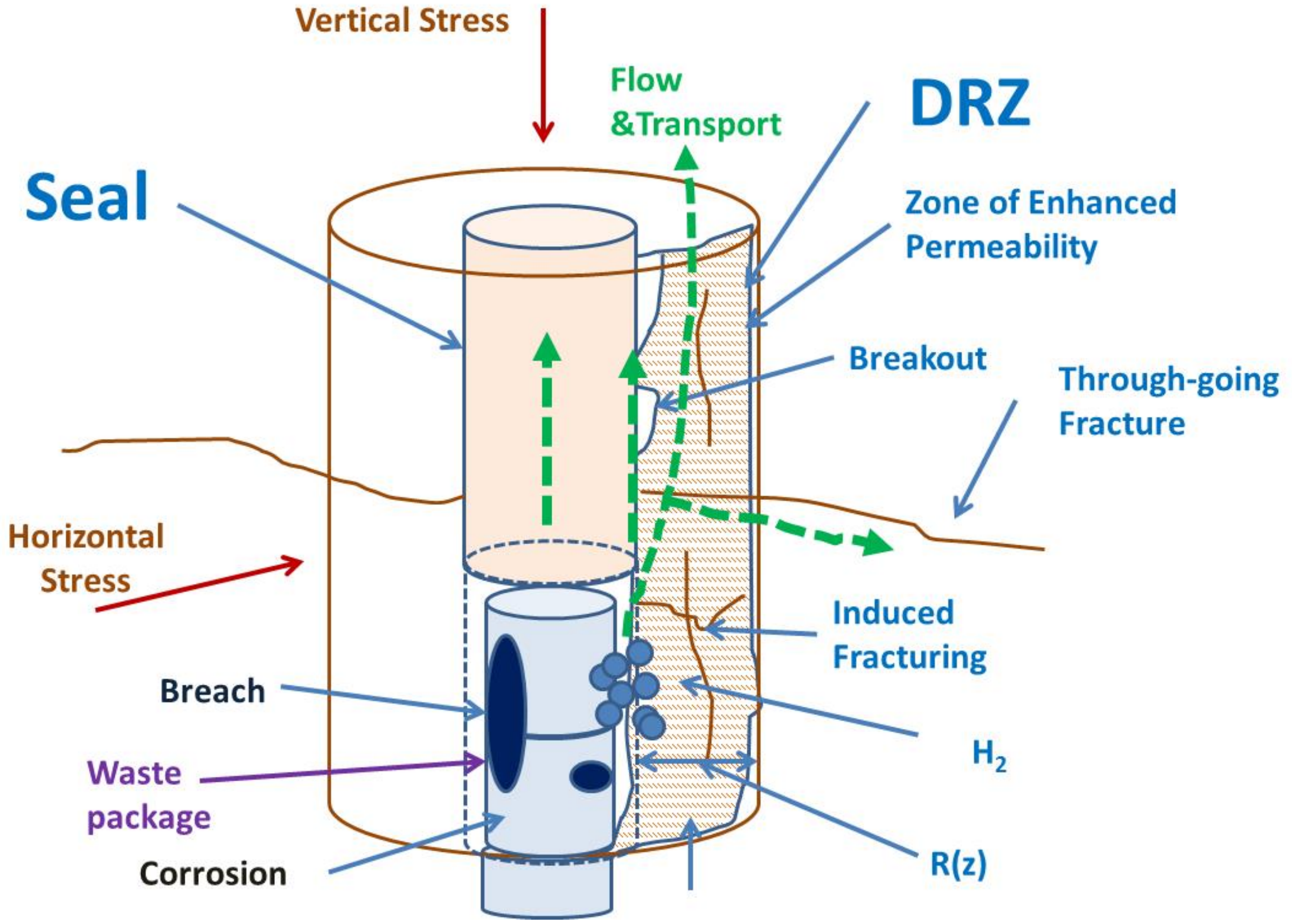

Figure 2. Schematic of the borehole environment (MacKinnon 2015).

Figure 2 illustrates several of the primary features and processes that are potentially important in the borehole environment. Although the schematic appears complex, the processes, e.g. borehole breakout, waste package breach, induced fracturing, won't all be happening simultaneously. Instead, they will be spread over 5 distinct time periods – drilling, emplacement, sealing, early post-closure (i.e., thermal pulse), and long-term post-closure. Therefore, these processes can be sometimes treated in isolation. Borehole breakouts and induced fracturing will happen during or shortly after drilling. The zone of enhanced permeability referred to as the disturbed rock zone (DRZ), will be established then. Waste packages are expected to remain intact during emplacement and sealing. After sealing, the geomechanical conditions of the rock will essentially be fixed. During the ensuing thermal pulse, old formation water will re-enter and re-equilibrate with the borehole, the waste packages will corrode, and, at some point, radionuclides will likely be released from the packages into the borehole. During the thermal pulse, temperatures will rise approximately 100$^{o}$C or less, above the ambient of ~ 100$^{o}$C, for a few decades. Ambient temperatures will prevail again after a few hundred years. The peak and duration of the thermal pulse depends upon the specific composition of the disposed waste. No credit is currently taken for waste package longevity in borehole performance assessments; instead the package is conservatively assumed to breach after emplacement and sealing. The use of corrosion-resistant waste package materials that would maintain package integrity for a few hundred years (i.e., through the early thermal pulse when upward fluid velocities would be greatest) would provide an additional barrier to delay radionuclide release; however, waste package longevity is not essential for long-term system performance.

Deep borehole disposal for the geologic isolation of nuclear waste was first evaluated by the US National Academy of Sciences (1957), O'Brien et al. (1979), and Woodward-Clyde (1983). Much of the recent research on deep boreholes has been led by the University of Sheffield (e.g. Gibb 1999; Gibb, McTaggart, Travis, Burley and Hesketh 2008; Gibb, Taylor and Bukarov 2008; Gibb, Travis and Hesketh 2012), MIT (Hoag 2006; Driscoll, Lester, Jensen, Arnold, Swift and Brady 2012; Bates, Salazar, Driscoll, Baglietto and Buongiorno 2014; Bates 2015) followed by Sandia National Laboratories (e.g. Brady et al. 2009; Arnold, Swift, Brady, Orrell and Freeze 2010; Arnold, Brady, Bauer, Herrick, Pye and Finger 2011). In 2012, the Blue Ribbon Commission on America's Nuclear Future (BRC) reviewed the prior research on deep borehole disposal, concluded that the concept may hold promise, and recommended further research, development, and demonstration to fully assess its potential. In 2013, consistent with BRC recommendations, the U.S. Department of Energy (DOE) proposed developing a research and development plan for deep borehole disposal. In 2014 the DOE began planning a deep borehole field test (DBFT) (SNL 2014). Site selection for the DBFT, based on an open request for proposal (RFP) process (DOE 2015 - DE-SOL-0008071), is scheduled for January 2016; drilling of a deep characterization borehole is expected to start by September, 2016. The DBFT will involve no radioactive waste, but is instead a science and engineering demonstration.

The goals of this chapter are to describe the state of the art of deep borehole disposal and to anticipate how the field of deep borehole disposal will evolve in the coming years. The sections below describe candidate wastes, deep borehole disposal facility siting considerations, and drilling, emplacement, and sealing methodologies. Two post-closure performance assessments for deep borehole disposal are summarized, both of which show minimal radionuclide releases from the crystalline basement: one for spent nuclear fuel (SNF) and one for high-level waste (HLW), specifically cesium and strontium (Cs/Sr) capsules, followed by a discussion of pre-closure operational safety. The aim and approaches of the DOE DBFT are then outlined, followed by long-term research that might improve confidence in the deep borehole disposal concept.

## Candidate Wastes

Volumes of different types of US nuclear wastes are shown in Figure 3. In 2014, DOE outlined a revised strategy for management and disposal of these wastes (DOE 2014) which recommends: disposal of commercial SNF and HLW, along with DOE-managed HLW and SNF with relatively higher heat output, in a geologic repository; disposal of DOE-managed HLW from defense activities and some thermally cooler DOE-managed SNF in a separate repository; and consideration for disposal of smaller DOE-managed waste forms in deep boreholes.

Commercial SNF, volumetrically the largest fraction of the US inventory, is not currently a candidate for deep borehole disposal in the US because, as described above, it is projected for a mined repository. However, pressurized water reactor (PWR) assemblies would fit in a borehole with a 17-in (43 cm) disposal zone diameter; a 2,000-m long disposal zone would accommodate about 400 singly stacked PWR assemblies. Hypothetically, the large US volume of commercial SNF would require between 700 and 950 boreholes (Arnold et al. 2011). For countries with limited volumes of SNF from reactors, the entire SNF and HLW inventory could fit in a small number of boreholes, and be more economically feasible than a mined geologic repository. Smaller DOE-managed waste forms that are candidate for deep borehole disposal in the US include (DOE 2014): Cs/Sr capsules from the Hanford Site; untreated calcine HLW; and salt wastes from electrometallurgical treatment of sodium-bonded fuels. The Cs/Sr capsules are of particular interest for deep borehole disposal because they are stored above ground in a large water pool and account for ~ a third of the radioactivity at the Hanford Site, but are volumetrically minor (even if vitrified as shown in Figure 3). If disposed directly, the entire Cs/Sr capsule inventory could fit in a single borehole with a 8.5-in (22 cm) diameter borehole and a 1,110-m long disposal zone. The motivation to put these wastes underground is that DOE post-Fukishima safety analyses identified the Cs/Sr capsule surface storage facility as having the highest catastrophic failure risk of any DOE facility.

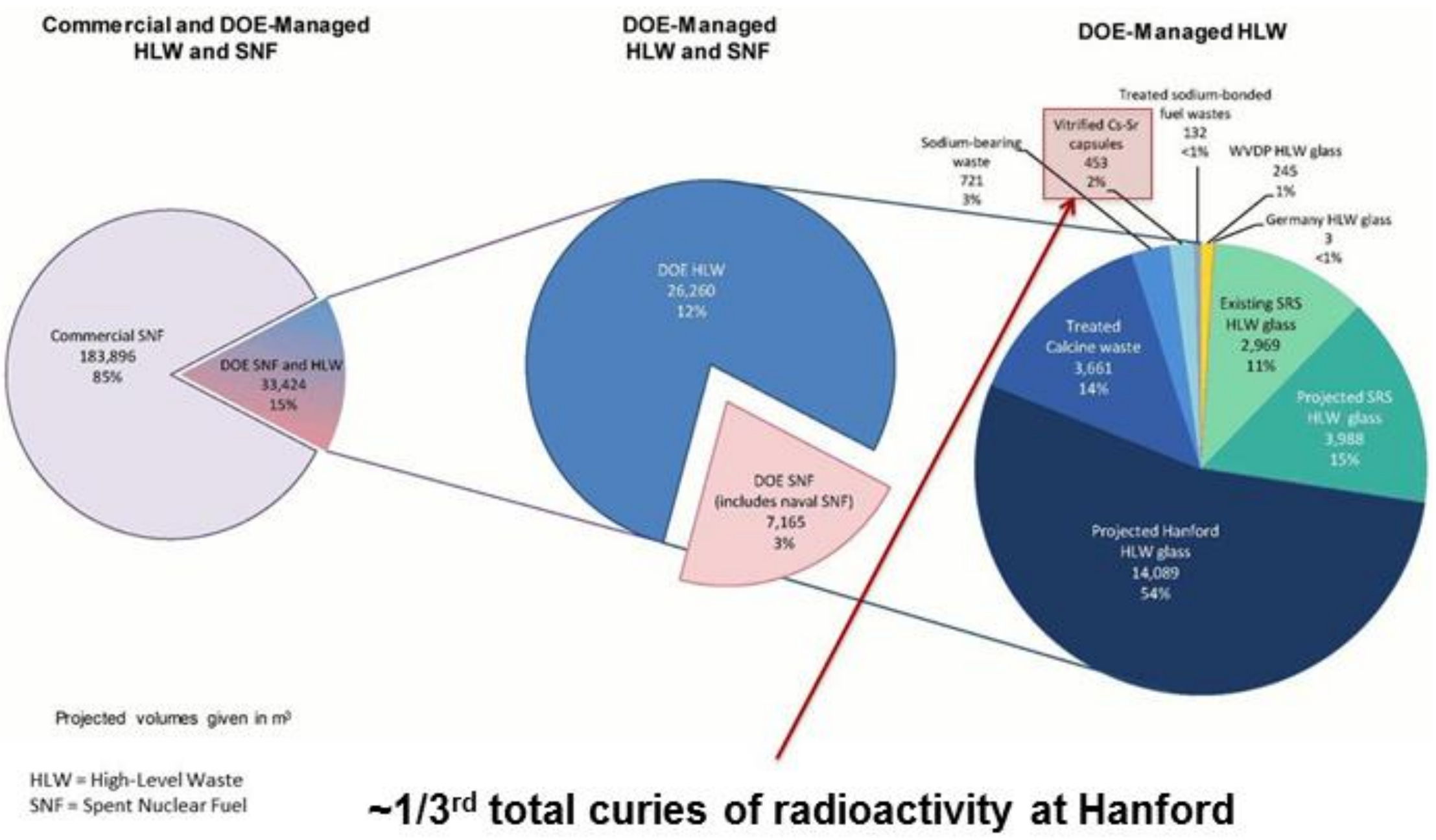

Figure 3. Radioactive waste volumes in the US (after Figure 1 of DOE 2014).

## Siting

Deep borehole disposal siting means maximizing the probability of successfully (i) drilling and completing a deep large-diameter borehole at a site with favorable geologic, hydrogeochemical, and geophysical conditions, (ii) building and maintaining the associated infrastructure, (iii) conducting surface handling, emplacement, and sealing operations, and (iv) demonstrating long-term post-closure safety. Overall, one of the important attractions of deep borehole is the potentially lower cost of site characterization compared to that of mined geologic repositories. Bates (2015) estimated the Yucca Mountain site characterization costs at $54/kgHM, a very large fraction of the total Yucca Mountain outlays.

Appropriate geologic sites for deep borehole disposal might be most readily identified by disqualifying sites with adverse characteristics, such as: upward vertical fluid potential gradients, economically exploitable natural resources, high-permeability connections from the waste disposal zone to the shallow subsurface, and significant probability of future seismic and/or volcanic activity. This process of elimination should leave sites sites that possess the following, more favorable characteristics.

- Depth to crystalline basement of less than 2,000 m – A depth less than 2,000 m allows for a 2,000 m disposal zone overlain by at least 1,000 m of seals within the crystalline basement.
- Crystalline basement geology that tends toward regionally more predictable structure and lithology. For example, plutonic or felsic intrusive rocks that tend to be less likely to have (a) major faults or shear zones, (b) well-connected fracture systems, or (c) recent tectonic activity or seismicity.

- Relatively low differential horizontal stress (low enough to allow borehole construction and utilization) – A large differential in horizontal stress at depth can lead to difficulties in drilling a vertical hole and maintaining borehole stability (e.g., borehole wall collapse and/or an enhanced disturbed rock zone around the borehole).
- Low seismicity – Seismic hazard could increase risk during drilling and emplacement. Seismic hazard is also a general indicator of tectonic activity, potential fault movement, and structural complexity.
- No volcanism – Quaternary-age faulting and volcanism is an indicator for potential future tectonic activity or volcanism.
- Low potential for deep circulation of meteoric ground water (e.g., low topographic relief and low hydraulic gradient) – Hydraulic gradients in the deep subsurface are generally related to regional variations in topography and can lead to the potential for upward flow in regional discharge areas. However, deep groundwater can be isolated and stagnant in some hydrogeologic settings, in spite of topographic effects. Long-term hydrologic stagnation is desirable.
- Favorable geochemical environment – High density, stratified salinity and geochemically-reducing conditions tend to reduce radionuclide mobility. Fluid compositions that are rock-dominated indicate long-term geologic controls in a relatively isolated fluid system.
- Low/normal geothermal gradient – Geothermal heat flux can lead to the potential for upward hydraulic gradients and is also related to the potential for geothermal drilling, a potential natural resource.
- No natural resource potential – Petroleum and mineral resources exploration and/or production could lead to human intrusion into the deep borehole and amplify the release of radionuclides to the overlying sediments. Proximity to oil and gas drilling activities can be beneficial during the borehole drilling and construction phase, decreasing some costs associated with transportation to remote sites.

Figure 4 shows areas of the continental US with basement depths less than 2 km, surface granitic rocks, Quaternary faults, recent volcanic activity, and recent seismic activity (ground motion > 0.2g).

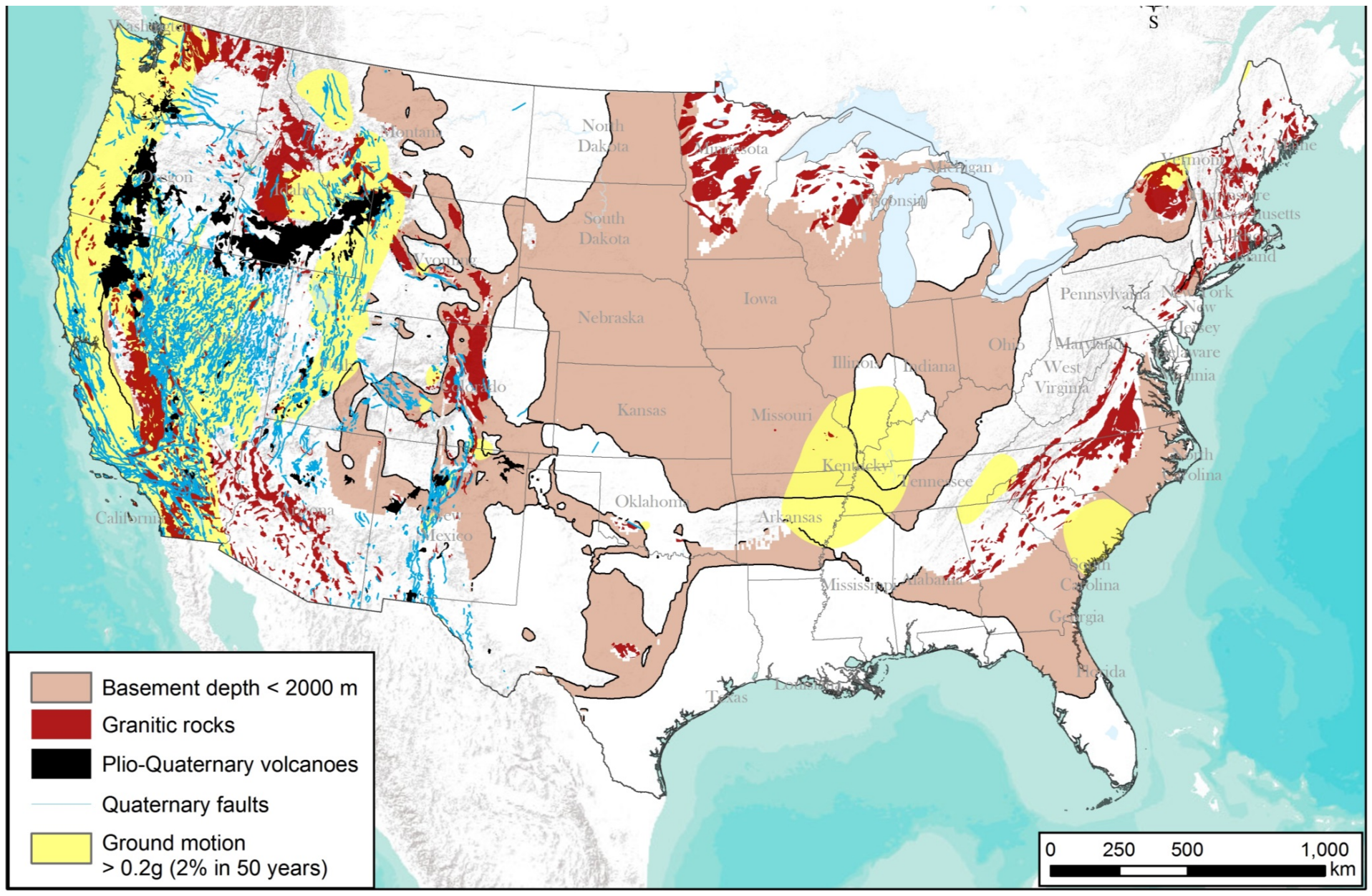

Figure 4. Basement depth, Quaternary faults, volcanic and seismic activity (from Arnold et al. 2013).

Figure 5 shows the areal extent of US oil and gas exploration and production activity.

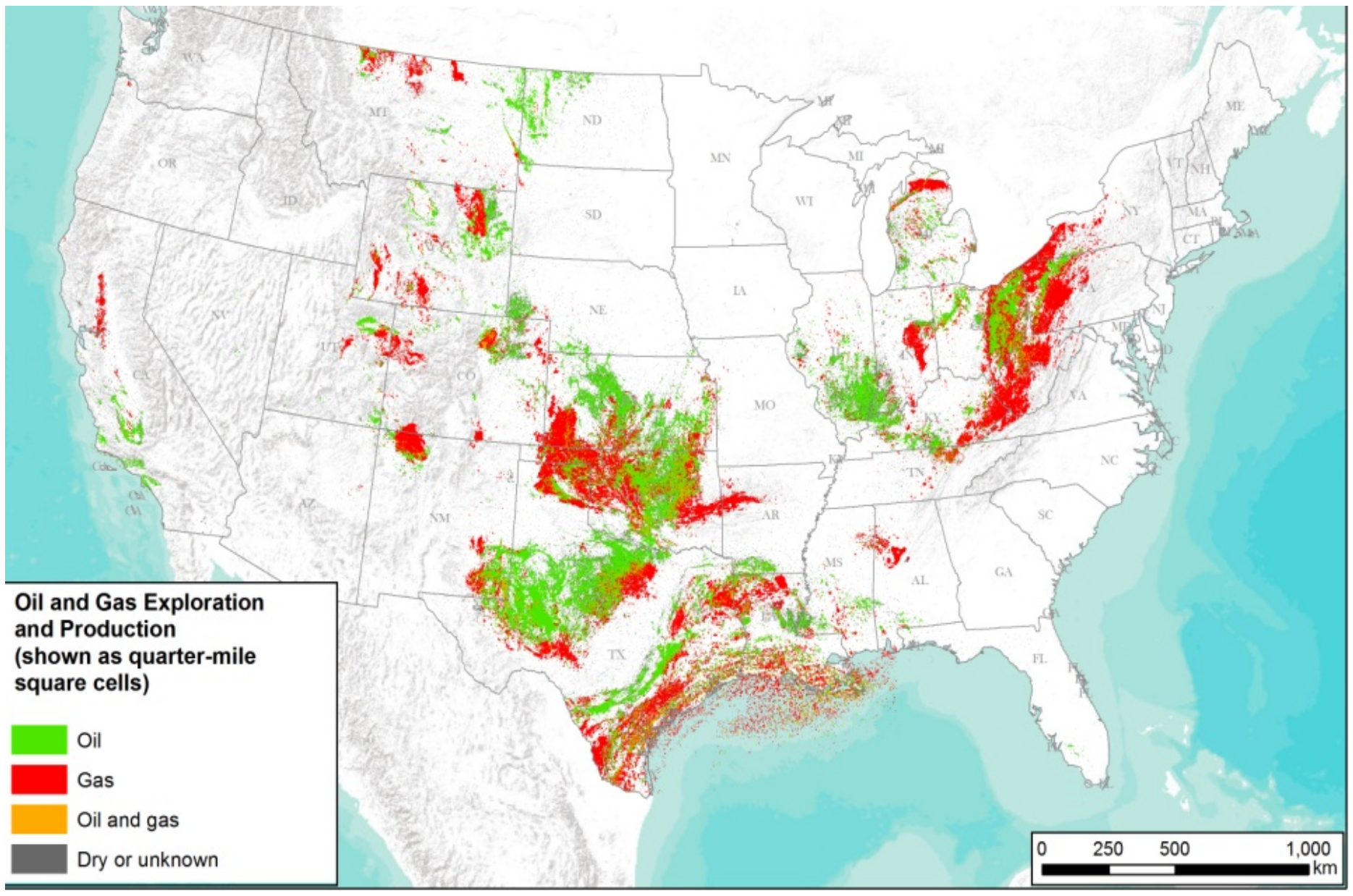

Figure 5. US oil and gas exploration and production activities (Source: USGS Digital Data Series, 069-Q).

Information like that presented in Figures 4 and 5, along with an analysis of heat flow, can be used to screen areas with characteristics favorable to deep borehole disposal. Note though that Figures 4 and 5 do not show fine-scale detail and are therefore are most useful for site-screening. Region-specific evaluations must be done to finally identify specific locations with appropriate characteristics for disposal. These evaluations would be done in conjunction with any process for consent-based siting that may be in place.

Once a site location is identified, several logistical factors must be worked out, including securing drilling contractors and support services, satisfying legal and regulatory requirements associated with drilling, construction of surface facilities for waste handling and emplacement, and assuring there is a sufficiently large surface site for drilling, surface handling, and emplacement operations. There should also be reasonable access to roadways and/or railways for transportation of waste and other materials.

## Drilling

Drilling technology has matured significantly since the Woodward-Clyde deep borehole design study in 1983. The advancements have been primarily associated with directional control, which is related to the boom in oil and gas drilling associated with horizontal wells. While deep borehole disposal is currently only being investigated in vertical boreholes, the same directional drilling technology can be used to maintain borehole straightness (i.e., dogleg-severity or maximum angular deviation across a specified distance) and verticality (i.e., borehole plumbness), even when the rock structure, fabric, or fractures would tend to cause the drill bit to deviate from vertical.

We broadly group relevant deep drilling methods by how drilling torque is applied to the drill bit, how directional control is maintained, and the type of drill bit.

Historically, drill rigs applied the torque to the drill bit through the drill pipe by an uppermost "kelly" section. The kelly is a piece of non-round cross-section drill pipe that is turned using a motor connected to a similarly shaped bushing fixed to a rotating table at the drill rig floor. The entire length of drill pipe is torqued to turn the drill bit at the bottom of the hole. Pipe is added to the bottom of the kelly section when advancing the hole.

More recently, top-drive motors have been utilized to turn the drill string. These involve the rotary motor being directly connected to the drill pipe at its top. The rotary motor assembly moves up and down the drill rig mast during drilling operations. While this is mechanically more complex than using a stationary kelly system, more control is allowed the drilling operator, including applying rotation while pulling up.

Downhole mud motors are a modern alternative method for applying torque to the drill bit. In these systems the drill pipe is not rotated; a positive displacement motor is part of the bottom of the drill string above the drill bit. Pumping mud down the drill string (i.e., direct circulation) then turns the pump, which translates into torque applied directly at the drill bit. Early directional drilling relied on this method, utilizing non-straight drill pipe (i.e., "bent subs", which could not be rotated) to achieve the bends required to perform horizontal drilling.

Both top-drive and kelly-drive systems can be configured to utilize reverse circulation, which pumps the drilling mud up the drill pipe, rather than down the drill pipe. This approach often results in more depth-specific cuttings retrieval than direct circulation, where mud circulates up the borehole annulus. Very large borehole diameters sometimes require reverse circulation to effectively remove cuttings, since mud flow velocities drop off as the annulus diameter increases (larger cuttings fall out of the mud when it slows down), while velocity in the drill pipe remains high. Reverse circulation is not compatible with some modern drilling approaches (e.g., downhole mud motors or hammer drilling) or would require specialized equipment.

For directional control, several different types of hybrid rotary steerable systems have recently emerged. These methods typically require the drill string to turn (via kelly or top drive), but have computerized active directional controls located at the bottom of the drill string above the drill bit. Current methods either dynamically apply a horizontal force to the drill pipe (i.e., pads dynamically push against the borehole wall to divert the bit a specific direction) a few meters above the drill bit, or dynamically bend the drill string during rotation to get the proper pointing of the drill bit. These rotary steerable systems can be much more expensive than downhole mud motors or more traditional drilling methods, but can maintain precise control of the straightness and verticality of the borehole through continuous surveying and downhole measurement while drilling. Downhole mud motors and multiple steerable system were used in the German KTB borehole which had excellent directional control to approximately 6 km depth (Bram et al. 1988), but the downhole electronics failed in the higher-than-expected temperatures encountered below that depth (Engeser 1995). Modern electronics in rotary steerable systems are now typcially tolerant of high temperatures, making this approach more feasible.

Drilling bits used in hard rock are typically rotary roller-cone type bits which have multiple rotating components covered in carbide buttons, which rotate and break up the rock at the bottom of the hole through compressive failure. Polycrystalline diamond compact (PDC) bits are a newer type of drilling bit developed for use in sedimentary rocks. These bits have no moving parts, and instead break the rock up through shear failure; cutter faces are dragged along the bottom of the borehole. PDC bits are much more expensive than roller-cone bits, but they have very high penetration rates and typically last much longer (requiring fewer trips out of the borehole for bit replacement). Some advanced PDC bits and hybrid roller-cone/PDC bits have recently been developed for drilling in hard rock, but there is less experience with crystalline rock, compared to the extensive recent experience with PDC bits in sedimentary rocks and the long history of using tricone bits in crystalline rock.

Hammer drilling is an alternative drilling method and drill bit type that conceptually replaces the downhole mud motor with a drilling-fluid activated drilling hammer. The hammer then compressively breaks up the rock at the bottom of the hole through rapid vertical up-and-down motion. Traditionally, most hammer drilling is done with air as a drilling fluid, but some experimental water-based hammer drilling methods are available. While hammer drilling can achieve very high penetration rates in hard rock, using air as drilling fluid is often undesirable at significant depth. It can be difficult to remove water that flows into the borehole with only air circulation, the compressibility of air and leakage of air from the joints in the drill string becomes significant in a very long drill string, and air-based drilling requires an underbalanced drilling approach that removes drilling fluid weight as a possible tool in managing the stability of the borehole.

Key criteria for selecting a suitable modern (i.e., directional-drilling capable) drilling rig in addition to borehole depth, diameter, and rock type include the expected weight of the drill string and the weight of casing/liner to be installed. Oil-field drilling rigs are available up to 4,000 horsepower size with lifting capacities up to 900 metric tons (Beswick 2008). Within the range of available land-based rigs, there are several rigs that are capable of drilling a large diameter borehole to 5 km in crystalline basement rock.

Top-drive rotary drilling in the crystalline basement would likely be performed using a hard-formation, tungsten-carbide insert, journal bearing, roller-cone bit. A downhole mud motor could be fitted with hybrid roller-cone/PDC bits. Deep borehole disposal should take advantage of recent advances in drilling and completion technology, but we should not be using experimental approaches, unless the consequences of failure for these approaches are acceptably low.

The choice of drilling method, and the selection of specific bits and operating parameters (rotary speed, bit weight, and mud hydraulics), will be driven by local drilling experience and rock characteristics at the site. Drilling in crystalline rock will be slow, with penetration rates possibly as low as 1 meter per hour. Hard crystalline basement rock will typically limit drill bit life. Frequent bit changes will increase the number of trips in and out of the borehole. Coupled with the large diameters, this means that drilling costs are somewhat uncertain. When drilling deep boreholes in hard rock, the amount of time spent tripping drilling and testing equipment in and out of the borehole (e.g., to change the drill bit, retrieve core samples, conduct a drill-stem test, or perform hydrofracture tests) can be a significant portion of the total time. This can be minimized by using longer drill pipe sections, longer-life drill bits including new hybrid types, alternative drilling methods, and wireline coring.

The fluid circulation system is composed of pumps, connections to the drill string, fluid recovery equipment, and surface equipment for fluid makeup and removal of cuttings. Depending on the drilling method the circulating fluid can be composed mostly of water, oil, or air. Its functions are to cool and lubricate the bit, lubricate the drill string, flush cuttings from the borehole, condition the hole to limit sloughing and lost circulation, and control downhole pressure. Drilling fluid or mud often has a significant impact on the cost of the borehole, particularly when the borehole has large diameter or lost circulation. The drilling fluid used in drilling the overburden section of the borehole will be selected to efficiently maintain a stable borehole across the overburden (e.g., water- or oil-based fluid with bentonite). Depending on the geology of the overburden, and the potential for clay sloughing or swelling, some sections of the hole may require oil-based fluid (e.g., for swelling clays) or brine (e.g., where evaporite minerals are present).

Cementing operations are important for ensuring the stability of casing strings and liners. Cementing may also be used to seal permeable zones and fractures during drilling, where lost circulation is encountered and other methods are not successful. Cement bond logs of cemented, cased intervals of the completed boreholes are used to confirm proper cement placement. Extended leak-off tests can be conducted at the bottom of cased intervals to verify cement performance.

## Emplacement

Although deep borehole waste emplacement operations are expected to be safe and accidents rare, the consequences of accidentally breaching a package during emplacement operations could be costly. Remediating a breached waste package downhole could involve decontaminating large pieces of drilling equipment, disposal of large volumes of radioactive drilling fluid, and possible worker dose. Accordingly, waste packages and systems for handling and emplacement must be designed with appropriate factors of safety that may exceed the safety factors typically assumed in oilfield applications.

Waste package performance requirements for deep borehole disposal are unique among alternative disposal concepts in that packages must withstand the bottom-hole hydrostatic pressure and stacking loads from packages emplaced one on top of another, while maintaining containment for a period of years until emplacement and sealing are completed. Packages can withstand hydrostatic pressure if they are robust, with multiple sealing elements. Alternatively, they can be less robust if filled with a fluid and equipped with pressure equalization (compliant elements such as pistons or sliding seals that maintain containment while transmitting volume). The robust, sealed approach could eliminate concerns with fluid interacting with waste, and eliminate the possibility of a fluid mobilizing solid waste in the event of accidental breach at the surface. Stacking loads on waste packages emplaced in a borehole can be limited by installing plugs in the borehole to bear the weight of additional packages.

Various methods for emplacement have been proposed: 1) lowering strings of waste packages on drill pipe; 2) stacking packages in a conveyance casing and lowering that on drill pipe; 3) lowering one or a few packages at a time on coiled tubing; 4) lowering one package at a time on an oilfield-type electric wireline; and 5) dropping packages one at a time for free fall to the bottom of the borehole. Each of these has potential advantages, and there is a wide range of likely cost. In general, the use of a workover rig and many tons of drill pipe to lower heavy strings of packages, or a heavy conveyance casing, greatly increases the risk of package breach if any part of the string is dropped in the borehole. By contrast, lowering packages a few at a time (and especially one at a time) using lighter wireline or coiled tubing equipment, lowers the potential energy released in the event of an accidental drop. The free-fall or “drop-in” method (Bates, Driscoll and Buongiorno 2011) depends on managing terminal sinking velocity and the force of impact, and on verifying the locations of packages once placed in the borehole. The wireline method and the drop-in method are actually similar in that both rely on sinking, while the wireline method provides real-time package status indications. With lighter equipment there is the possibility of using downhole impact limiters to further limit the probability of breach. A comparison of safety considerations between drill-pipe and wireline methods is discussed later in this chapter.

One of the hazards of emplacement is getting one or more waste packages stuck in the borehole, particularly if stuck above the intended disposal zone. A straightforward approach to mitigate this hazard is the use of guidance casing of constant size, from the surface to total depth of the borehole. The only complications have to do with cementing the disposal zone casing so that it does not bear the weight of all waste packages in column, and for packages (or conveyance casing) emplaced on drill pipe, a positive connection is maintained at all times so that if the string gets stuck, the necessary connection to pull it free is already established. A similar situation exists for wireline and coiled tubing but with less available pulling force. In this case, if pulling force is insufficient, then stuck packages can be released on command, and a drill rig brought in to connect and pull out with drill pipe. Ultimately, a package stuck in the disposal zone could be left alone, and a package stuck above the disposal zone could be removed by pulling the guidance casing. This last resort can be facilitated by hanging separate sections of guidance casing in the disposal zone and the interval above. The DBFT demonstration discussed below will use guidance casing hung in two intervals (i.e., a liner in the disposal zone and a tieback above) for demonstrating the emplacement and retrieval of test waste packages.

## Seals

Borehole seals are important for limiting vertical fluid movement during the thermal pulse, the duration of which is largely determined by decay of relatively short-lived fission products in the first few hundred years after emplacement. Seals performance for longer periods of time would be desirable, though the absence of a driving force for vertical fluid flow after the passing of the thermal pulse makes long-term seal performance less critical. Traditional candidates for borehole seal materials are bentonite and cement. Bentonite is attractive because it expands in contact with water and has a high surface area; for these reasons it has routinely been used to seal oil and gas and geothermal boreholes. Likewise, bentonite has been extensively studied as an engineered barrier in mined repositories. New finer-grained cements are being developed for more effective sealing in fractured rocks. Rock welds and thermite are recently developed concepts. Rock welding uses a resistance heater to melt crushed granite into a “weld” similar in makeup to the native crystalline rock. Thermite plugs form rapidly upon ignition of an Al-Fe metal-oxide charge (Lowry and Dunn 2014).

An effective seal should have a low permeability, bind effectively to the surrounding DRZ, be free of fractures and void spaces, be relatively straightforward to emplace, be resistant to chemical alteration which might affect permeability, and perform for at least several hundred years. Evaluating the long-term effectiveness of sealing materials for deep borehole disposal involves using industry experience, pressure tests, and/or laboratory measurements to estimate likely initial permeabilities of seals and the DRZ, and projecting permeability changes over long periods of time using lab and/or theoretical calculations to estimate chemical and physical alteration impacts on permeabilities.

**Bentonite**

Bentonite swelling and sealing depends upon the salinity and chemical makeup of the water that contacts the bentonite. Bentonite swelling increases, hence permeability decreases, as ionic strength and the ratio of divalent to monovalent cations in solution decrease. Water-saturated bentonite tends to have a density of 1700 – 2100 kg/m$^3$ and a permeability of less than $10^{-11}$ m/s. Brines at the bottom of the borehole will have high ionic strengths and high divalent cation levels, leading to less bentonite swelling; fluids above the waste emplacement zone will be more dilute, leading to more bentonite swelling. Bentonites near cement may be subjected to high $Ca^{+2}$ levels and high pH; and bentonites near degrading steel may see high $Fe^{+2}$ and $Ni^{+2}$ concentrations.

Bentonite swelling decreases with temperature and over long periods of time, as the bentonite transforms to non-swelling illite with attendant silica cementation:

$$\text{Bentonite} + Al(OH)_3 + K^+ \leftrightarrow \text{Iliite} + SiO_2$$

High $Na^+$ activity and restricted $K^+$ supply slow illitization, as does restricted Al supply. While it is generally assumed that bentonite performance decreases above 120°C, Wersin et al. (2007) have shown that the illitization/cementation probably occurs only above 150°C and that, even then, bentonites are likely to maintain much of their sealing ability. Recall that borehole temperatures will exceed 150°C for only short periods of time.

One threat to bentonite seal performance is physical erosion by flowing water present in transmissive zones. Bentonite erosion depends on both the groundwater velocity and the chemistry of the water. Fresh water deflocculates bentonite and favors erosion. High TDS, high Ca brines would work against erosion. The TDS and Na/Ca ranges of solutions needed to deflocculate bentonite are reasonably well understood from the oil industry (e.g. Scheuermann and Bergerson 1990).

How to reliably deliver compacted, dehydrated clay down a fluid-filled borehole to an open uncased sealing interval is a subject of research. Some options are oil-based mixtures, highly compacted blocks wrapped in reactive membrane material, and pumped slurries.

**Cement**

Because cement is thermodynamically unstable in subsurface environments, it will transform to a more stable assemblage over time – possibly with a performance-reducing decrease in volume and permeability. Most cements experience a slight volume loss upon hardening. Leachate from cements in the borehole might decrease the performance of bentonite by supplying $Ca^{+2}$ and raising the pH. High $Ca^{+2}$ decreases the swelling pressure; high pH dissolves the bentonite. The overall effect on seals performance depends primarily upon the relative proportions of leachate and bentonite that react, the $Ca^{+2}$ and pH of the leachate, and the temperature. Low pH cements are being considered as a way to minimize leachate-bentonite reactions. Low pH cements are pozzolans, high silica cements developed by the Romans for use in water-rich settings. Expanding cements with CaO or MgO additives (API Class G cements) are often used in deep drilling. Adding bentonite to cement decreases the strength of the cement and the self-sealing ability of the bentonite. Borehole cement will face corrosion through carbonation and, possibly, sulfate attack which result in precipitation of, respectively, calcium carbonate and calcium sulfate in the near-surface cement matrix. A key uncertainty is the chemistry of the water contacting the cement, in particular its $HCO_3^-$ and $SO_4^{-2}$ levels, and pH.

**Rock Welding**

Rock welding involves partially melting crushed granite backfill and the granitic wall rock with a downhole electric heater. The melt then recrystallizes to a holocrystalline rock identical to, and continuous with, the host rock in almost all its properties except grain size. In theory, the rock welds are calculated to be large enough to seal the hole and locally eliminate the DRZ. Some small scale tests have been done in the past to evaluate melting properties of rock welds (e.g.

Gibb, Taylor and Bukarov 2008), but a greater effort is needed if a field-scale demonstration is to be done.

Industry measures the annular sealing of cased boreholes through cement bond logs and pressure testing. The effectiveness of cement plugs are likewise measured with pressure tests. There are currently no logging tools or techniques that 'see' into plugs using acoustic, resistivity, or density techniques (Radioactive Waste Managmement Directorate 2014). Nor are there long-term tests designed specifically to evaluate the long-term performance of cement-bentonite seals (Radioactive Waste Managmement Directorate 2014). One example of the type of analysis desired is the borehole cement analysis from the WIPP. The basis for cement longevity at the Waste Isolation Pilot Plant (WIPP) was determined for borehole plugs in experiments conducted by Thompson et al. (1996) who found that plug failure occurred when the cement matrix was measurably altered. Thompson et al. (1996) concluded that in a 3-plug borehole design, deeper casing corrosion would be less severe than upper sections and that deeper plugs (e.g. emplaced below 4 km) would not fail for approximately 5,000 years.

## Safety Analysis of Borehole Disposal of Spent Fuel

Licensing of any nuclear waste disposal site relies on numerical model predictions of post-closure performance. Performance analysis calculations done to date for deep borehole disposal point to very low releases. The first of these (Brady, Arnold, Freeze, Swift, Bauer, Kanney, Rechard and Stein 2009) considered disposal of commercial spent nuclear fuel and assumed:

- 400 Pressurized water reactor (PWR) assemblies, ~150 metric tons heavy metal (MTHM), vertically stacked down the 2 km length of the bottom of the waste disposal zone.
- Thermally driven hydrologic flow from the top of the waste disposal zone upward through 1,000 m of a bentonite sealed borehole with a specific discharge of 0.017 m/yr for 200 years.
- Pumping of borehole water from the location 1,000 m above the top of the waste disposal zone to the surface via a withdrawal well. No credit was taken for sorption or decay along the saturated zone transport pathway from the borehole to the withdrawal well.
- A dilution factor of $3.16\times10^7$ to account for mixing of borehole water with water in an existing aquifer before it would be captured by the withdrawal well assumed to supply 1,000 people.
- A transport time of 8,000 years was applied to account for the time taken for the bulk of the dissolved radionuclide mass to be captured by the withdrawal well.

Radionuclide transport up the borehole was calculated with a 1-dimensional analytical solution to the advection-dispersion equation. Radionuclide transport up the borehole from the waste disposal zone occurs for 200 years, corresponding to the duration of the thermally driven flow. Subsequent to the thermal period, ambient conditions are not expected to provide any upward gradient, and advective radionuclide transport was assumed to cease.

The source concentration at the top of the waste disposal zone was determined by (a) calculating a maximum potential concentration based on dissolving the entire initial mass inventory in a PWR into the void volume (i.e., the potential volume of water) of a waste package, and (b) selecting the lower of the maximum potential concentration and independently calculated solubility limits as the source concentration.

The only radionuclide with a calculated non-zero concentration 1,000 m above the waste disposal zone in the sealed borehole is $^{129}I$. The non-zero $^{129}I$ concentration ($5.3 \times 10^{-8}$ mg/L) represents the leading edge of the dispersive transport front. However, the center of mass never reaches the top of the 1,000 m sealed section of the borehole because there is effectively no further movement after the few hundred year thermal pulse due to the relative slowness of diffusion. Subsequent diffusive transport to the hypothetical withdrawal well decreases the $^{129}I$ dose further. The peak dose from the withdrawal well occurs at 8,200 years and is exceedingly small -$1.4 \times 10^{-10}$ mrem/yr and is solely from $^{129}I$. For comparison, the Yucca Mountain standard was 15 mrem/yr for the first 10,000 years, and 100 mrem/yr from peak dose to 1 million years.

These preliminary results are based on several bounding and conservative assumptions, such as: all waste is assumed to instantly degrade and dissolve inside the waste packages; the waste packages are assumed to maintain structural integrity during surface handling and emplacement, but are assumed to be degraded immediately after sealing; and no credit is taken for sorption or decay along the saturated zone transport pathway from the sealed borehole to the withdrawal well. Lastly, isotopic dilution by non-radioactive iodine in the groundwater could lower $^{129}I$ levels by a factor of up to > 100 (Bates 2015). More refined and physically realistic performance assessments will likely indicate lower doses, or later peak doses, or both.

## Safety Analysis of Borehole Disposal of Cs/Sr

Currently 1,936 Cs and Sr capsules are stored underwater at the Hanford Waste Encapsulation and Storage Facility. The capsules are less than 0.09 m (3.5 in) in diameter and are obvious candidates for deep borehole disposal. The capsules contain primarily short-lived $^{90}Sr$ and $^{137}Cs$, and long-lived $^{135}Cs$. Figure 6 shows a schematic of a borehole drilled to a depth of 5,000 m into crystalline basement rock, with a bottom-hole diameter of 0.22 m (8.5 in) with waste packages containing the Cs/Sr capsules emplaced in the lower disposal zone portion of the borehole, between 3,700 m and 5,000 m depth. Sealing and plugging the upper portion of the borehole is done with alternating layers of bentonite clay, cement, and cement/crushed rock backfill.

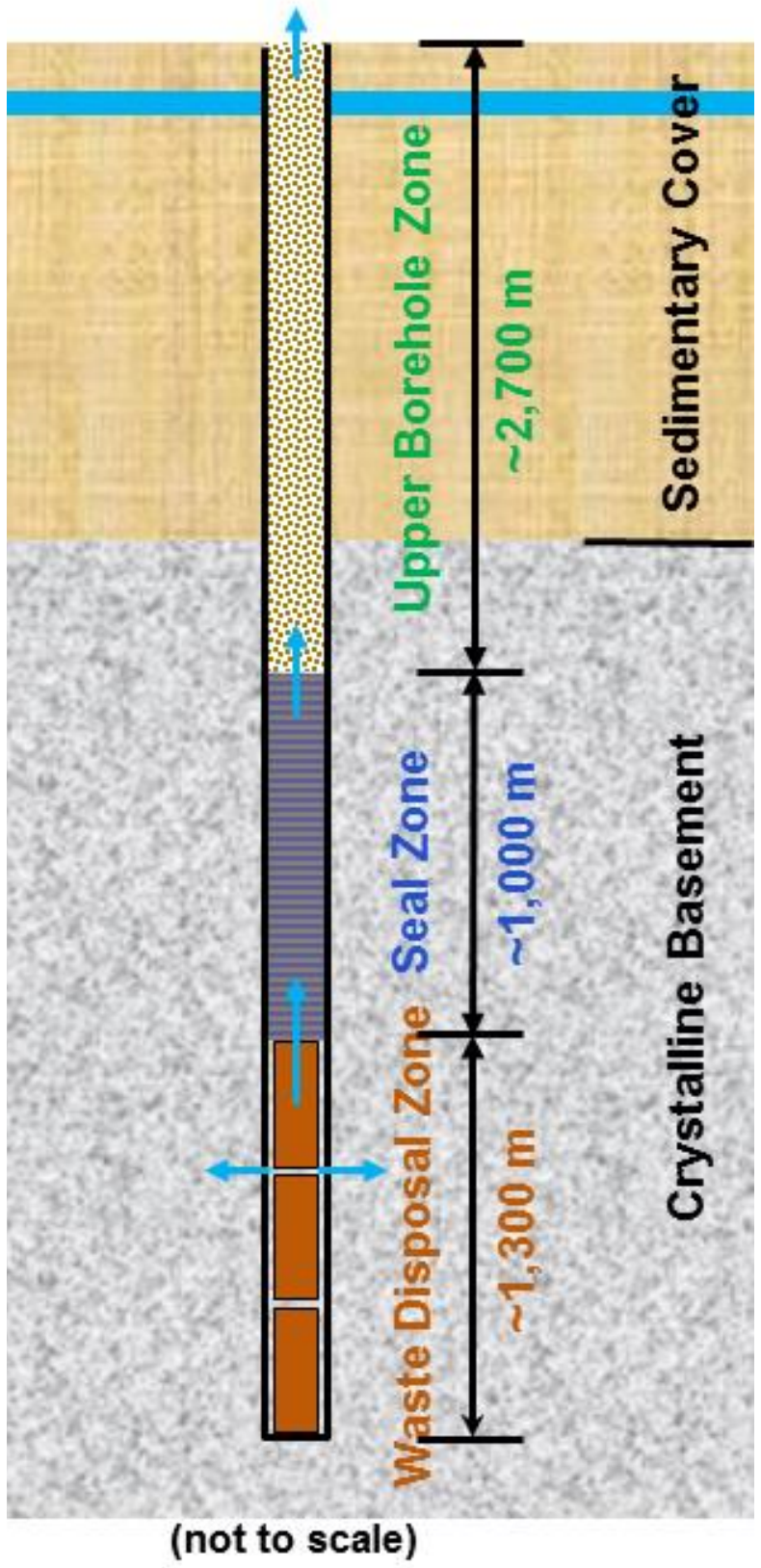

Figure 6. Baseline deep borehole disposal concept for Cs/Sr capsules.

Each waste package would contain two capsules end-to-end with a total waste package length of 1.08 m and an outside diameter of 0.11 m (4.5 in). In this baseline design, the 968 waste packages containing the Cs/Sr capsules would fit in a 1,300 m long disposal zone (this length includes spacing between waste packages) with a 0.22 m (8.5 in) diameter borehole. The baseline undisturbed scenario includes:

- 968 waste packages containing 1,335 Cs capsules and 601 Sr capsules in a 1,300 m waste disposal zone. The waste packages are assumed to maintain structural integrity during surface handling and emplacement, but are assumed to be degraded immediately after sealing.
- The 1,000 m seal zone has the bulk permeability and porosity of bentonite clay.
- The crystalline basement rock is assumed to have low bulk permeability and porosity; bulk permeability decreases with depth. Salinity and fluid density increase with depth.
- The DRZ around the borehole is assumed to have a permeability that is 10 times higher than the intact basement rock.
- The upper borehole zone has a bulk permeability and porosity of crushed rock backfill.

The low-permeability and low thermal conductivity of the surrounding crystalline host rock focus upward flow from the early thermal period through the borehole seals and/or the DRZ. The peak vertical groundwater flux (darcy velocity) through the seals/DRZ is about 0.01 m/yr for about 100 years. This corresponds to a pore velocity of about 1 m/yr, and a center-of-mass advective distance of about 100 m. The region of advective movement is only a small portion of the 1,000 m seal zone.

Following the approximately 100-year period of peak thermal perturbation, subsequent radionuclide transport to the biosphere is predominately by diffusion up the borehole seal and DRZ. At this time, most of the short–lived $^{90}Sr$ and $^{137}Cs$ have decayed away, leaving just a small mass of $^{135}Cs$ to contribute to longer-term dose. The baseline scenario results suggest that doses are quite low, even without any performance credit from the waste forms or waste packages. The dose is dominated by $^{135}Cs$. Peak dose occurs roughly 2 million years after emplacement and is less than $10^{-8}$ mrem/yr.

## Pre-Closure Safety

While the analyses above emphasize post-closure safety, pre-closure safety is also an important factor for both traditional geologic disposal and deep boreholes. Pre-closure safety considers potential hazards associated with waste package surface handling and downhole emplacement activities, which would require radiation shielding and/or remote handling operations; hazards include worker occupational safety, worker dose, and the potential for operational failures (e.g., waste packages stuck in a borehole above the disposal zone). Borehole pre-closure safety goals include:

- Borehole and casing that can be emplaced at the desired depth.
- Waste packages that don't leak during loading, transportation, handling, emplacement, and sealing of the borehole.
- Safe handling and emplacement of the waste packages.
- Not getting waste packages stuck in the borehole.

A preliminary hazard analysis of wireline emplacement of 400 waste packages in a prototypical deep borehole identified four top events affecting waste packages in the hole (Figure 7): 1) package drops from the top; 2) package drops during the trip in; 3) one or more packages getting stuck in the borehole during a trip in; and 4) drill pipe or wireline and tools dropped during a trip out. Each of these top events was associated with a probability of a resulting waste package breach. For events involving one or more stuck packages, the recovery operation (fishing) was also assigned a probability of success, and of package breach resulting from unsuccessful fishing. The purpose of this study was to evaluate the risks associated with different emplacement methods, with the intention of selecting one for demonstration in the DBFT (discussed below).

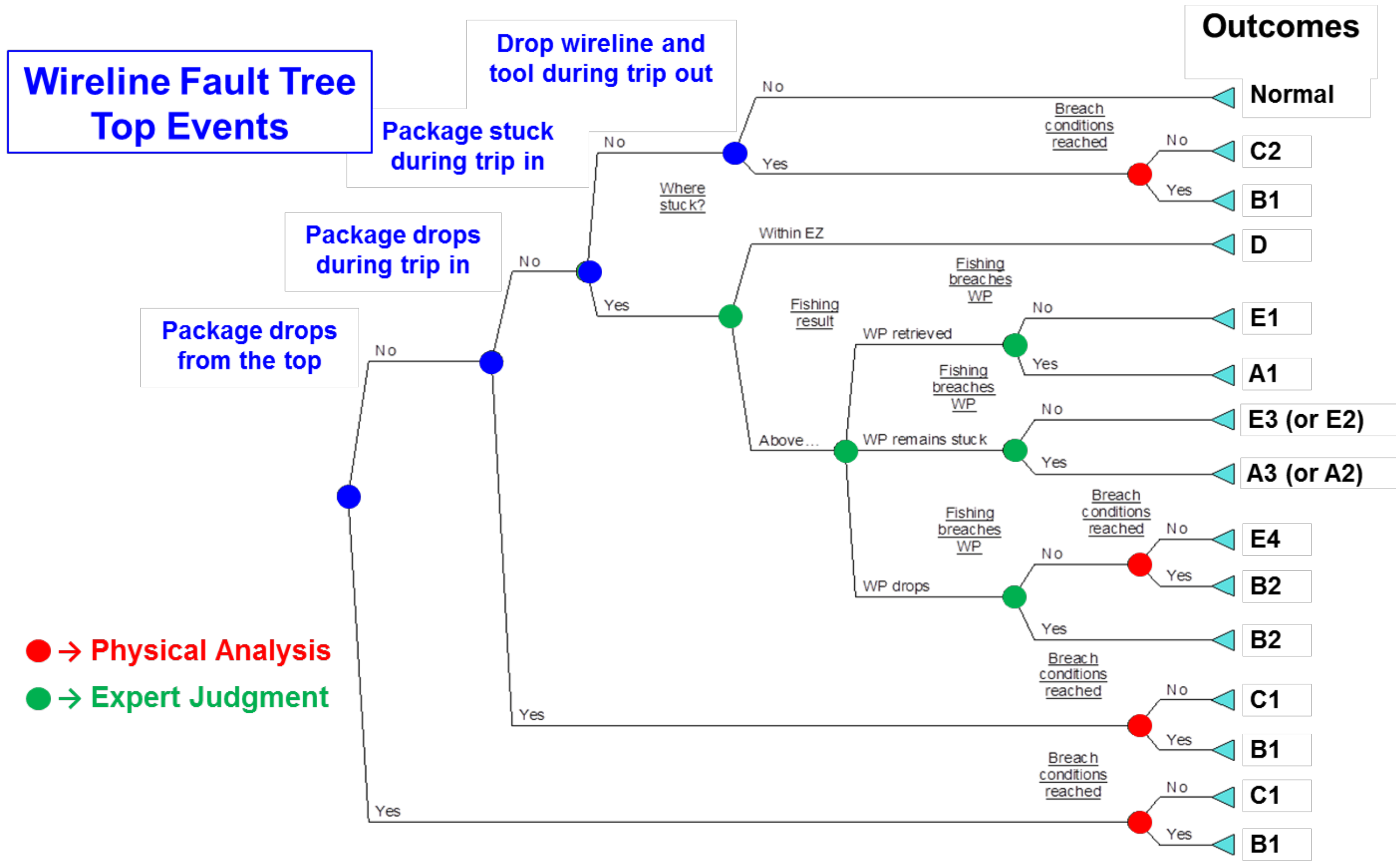

Figure 7. Event tree for wireline emplacement (from SNL 2015).

The study also considered consequences for the possible outcomes from the event tree for each alternative emplacement method (the wireline method is represented in Figure 7). Consequences included additional rig time for remediation, contaminated equipment, disposition of contaminated drilling fluid, pipe, and casing, and partial loss of disposal capacity in the borehole. Occupational hazards and worker radiological dose are also important but were not included explicitly in the study because industrial experience shows that they can be managed to acceptable levels, so they would not discriminate between the emplacement method choices.

Results from the study indicated that the probability of incident-free wireline emplacement of 400 waste packages is 96.81% and the probability of a radiation release (i.e., a waste package breach) is $1.29\times10^{-4}$. The probability of a waste package breach and radiation release during operations was estimated to be 55 times less for wireline operations, compared to drill-string operations. This was mainly because dropping a single package (with impact limiter) has little or no potential to breach a package, whereas dropping a string weighing 100 tons or more has a high likelihood of breaching a package on impact.

Similar analyses might be done to assess transportation safety, surface handling, worker exposure, and the effects of external events such as flooding, extreme weather, seismicity, and sabotage. Field experience at an actual borehole (e.g., the DOE DBFT) is expected to provide valuable input to future pre-closure safety analysis.

## Deep Borehole Field Test

The DOE DBFT, as planned, involves drilling two 5 km deep boreholes: an initial smaller-diameter Characterization Borehole for hydrogeological, geophysical, and geochemical investigations; and a subsequent larger-diameter Field Test Borehole for demonstrating surface handling, emplacement, and retrieval with surrogate test packages (the DBFT will not involve radioactive waste). The DBFT is a science and engineering demonstration to evaluate the safety and feasibility of siting, characterization, surface operations, and package emplacement activities as they relate to any future deep borehole disposal facility. Site selection for the DBFT, based on an open request for proposal (RFP) process, is scheduled for January 2016.

### Characterization Borehole

Once the site is selected, and permitting completed, a Characterization Borehole (CB) will be drilled to 5 km depth. The CB will be 0.22 m (8.5 in) in diameter, and drillable with existing technology. The overlying sedimentary section will be drilled and cased with minimal testing. Drilling and characterization of the crystalline basement will include: ~ 5% coring of the crystalline length; and testing and sampling after borehole completion using a packer tool via a work-over rig. Characterization activities in the crystalline basement will focus on measurements and samples that are important for evaluating the long-term isolation capability of the deep borehole disposal concept; therefore, drilling of the CB will be done with the aim of maximizing collection of usable samples.

Recall that establishing the relative age of waters in the crystalline basement is a key task of borehole site characterization. High salinity suggests long reaction with the rock, but is not a complete indicator of great age. Isotopic tracers provide a more comprehensive picture of groundwater age and provenance and will be utilized where possible in the characterization borehole.

Hydrogeologic testing in the characterization borehole will include measurement of static formation pressure and permeability/compressibility – pumping and sampling in high permeability strata, pulse testing in low permeability strata. Vertical dipole testing will be done to understand transport pathways. Hydraulic fracturing testing will be done to quantify subsurface stresses.

### Field Test Borehole

The Field Test Borehole (FTB), planned to have a 0.43 m (17 in) in diameter, will push the envelope of deep drilling technology. Figure 8 shows the design of the FTB. Planned activities in the FTB will evaluate the feasibility of: safely drilling a large diameter borehole to 5 km in crystalline rock, successful emplacement and retrieval of test packages, and demonstration of surface handling operations. These engineering demonstration activities will focus on providing data to evaluate operational safety and effective engineering solutions.

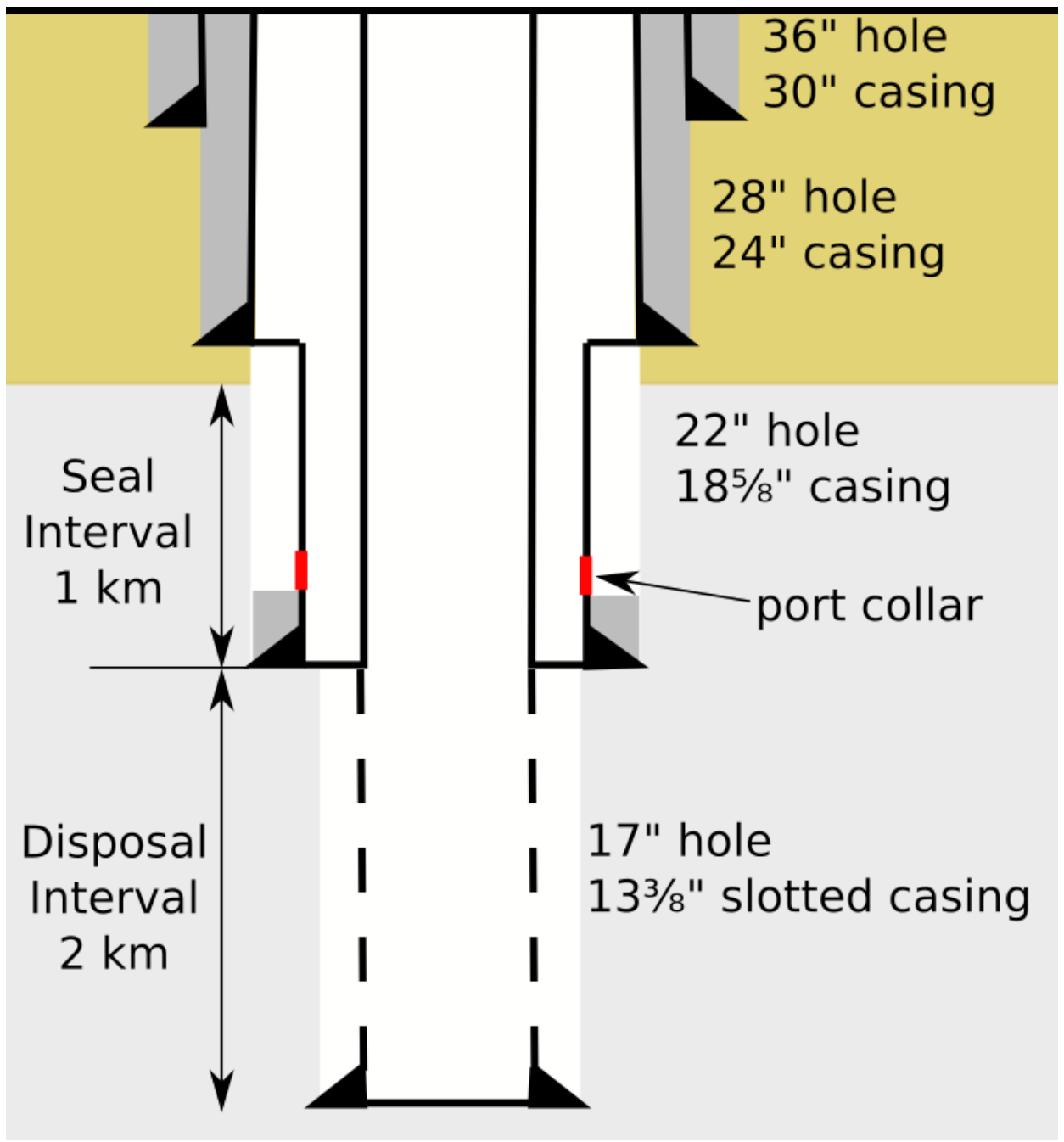

Figure 8. Field Test Borehole design.

A lower priority of the DBFT is borehole sealing. No field testing of borehole seals or sealing methods are planned during the DBFT. Instead, borehole sealing materials and emplacement configurations will be examined in parallel with DBFT field operations, starting from the reference seal design in Arnold et al. (2011). Key components to the DBFT seals effort will include: experimental analysis of bentonite alteration and steel corrosion under borehole conditions; examination of the seal-DRZ interface; consideration of newly developed sealing approaches including novel cements and thermite plugs, and; non-traditional approaches to borehole sealing such as rock-welding (e.g. Gibb et al. 2008).

## Conclusions

Deep borehole disposal is close to technically feasible today. Characterization, drilling, and emplacement of waste at a site could probably be done within 5 years, or sooner. And it would be cheaper than disposal in a mined repository while being just as safe.

## ACKNOWLEDGEMENTS

Sandia National Laboratories is a multi-program laboratory managed and operated by Sandia Corporation, a wholly owned subsidiary of Lockheed Martin Corporation, for the U.S. Department of Energy's National Nuclear Security Administration under contract DE-AC04-94AL85000. This work is supported by DOE Office of Nuclear Energy, Office of Used Nuclear Fuel Disposition. We greatly appreciate all the help and support we have received over the years from Prof. Mike Driscoll at MIT and Prof. Fergus Gibb at the University of Sheffield.